\documentclass[twocolumn]{autart}

\usepackage[longnamesfirst]{natbib}
\usepackage{amsmath}
\usepackage{amssymb}	% mathbb
\usepackage{mathtools}	% mathbb
\usepackage{color}
\usepackage{graphicx}	% graphics
\usepackage{lmodern}	% fontsize
\usepackage{url}

\newcommand{\R}{\ensuremath{{\mathbb R}}}
\newcommand{\Z}{\ensuremath{{\mathbb Z}}}

\newcommand{\XX}{{\mathcal X}}

\newcommand{\BB}{{\mathcal B}}
\newcommand{\AAA}{{\mathcal A}}

\newtheorem{thm1}{\bf Theorem}

\newtheorem{assmpt1}{\bf Assumption}

\newtheorem{clm1}{\it Claim}

\newenvironment{Assumption}{\begin{assmpt1}}{\hfill$\Diamond$\end{assmpt1}}
\newenvironment{theorem}{\begin{thm1}}{\hfill$\Diamond$\end{thm1}}

\definecolor{auburn}{rgb}{0.43, 0.21, 0.1}

\allowdisplaybreaks[4]

\begin{document}
	\begin{frontmatter}
		
		\title{Data-driven Output-feedback Predictive Control: \\ Unknown Plant's Order and Measurement Noise\thanksref{footnoteinfo}}
		
		\thanks[footnoteinfo]{This work was supported by the National Research Foundation of Korean(NRF)
grant funded by the Korea government(MIST) (No. 2020R1F1A1069426), and by AI based Flight Control Research Laboratory funded by Defense Acquisition Program Administration under Grant UD200045CD.}

		\author[SSU]{Nam H. Jo}\ead{nhjo@ssu.ac.kr}%
		\ \ and \
		\author[Seoul]{Hyungbo Shim}\ead{hshim@snu.ac.kr}             % e-mail address 

		\address[SSU]{Department of Electrical Engineering, Soongsil University, Seoul, Korea}
		
		\address[Seoul]{ASRI, Department of Electrical and Computer Engineering, Seoul National University, Seoul, Korea}

		\begin{keyword}                           % Five to ten keywords,  
		data-driven control, predictive control, uncertain systems, unknown order, Moore-Penrose inverse
		\end{keyword}                             % keyword list or with the 

		\begin{abstract}
The aim of this paper is to propose a new data-driven control scheme for multi-input-multi-output linear time-invariant systems whose system model are completely unknown.
Using a non-minimal input-output realization, the proposed method can be applied to the case where the system order is unknown, provided that its upper bound is known.
A workaround against measurement noise is proposed and it is shown through simulation study that the proposed method is superior to the conventional methods when dealing with input/output data corrupted by measurement noise.
		\end{abstract}
		
	\end{frontmatter}

%%%%%%%%%%%%%%%%%%%%%%%%%%%%%%%%%%%%%%%%%%%%%%%%%%%%%%%%%%%%%%%%%%%
\section{Introduction}
%%%%%%%%%%%%%%%%%%%%%%%%%%%%%%%%%%%%%%%%%%%%%%%%%%%%%%%%%%%%%%%%%%%%

Obtaining a mathematical model is the first step for model-based control designs, which has been however a difficulty in some applications.
This has motivated the study of model-free, data-driven control methods, and recently a method called Data-{\bf e}nabl{\bf e}d Predictive Control (abbreviated by DeePC) is presented by \cite{Coulson2019}.
The root of DeePC is the classical Model Predictive Control (MPC).
By noting that, at each sampling time, MPC finds an optimal control sequence for a finite time interval by evaluating a cost function based on the output sequence generated by a mathematical system model, DeePC simply replaces the model-based output sequence with a linear combination of the output data which are measured and stored from the previous experiments.
This replacement is justified by the celebrated {\em behavioral approach} by \cite{Willems2005}.

While DeePC has been successfully applied to several practices, some limitations are found from a few examples.
A limitation arises when an unstable system is the target of the control and the optimization horizon for MPC is not short.
In this case, the length of output data is not short, and hence, exponentially growing output data tend to cause numerical errors in optimization.
Another limitation is that, when the output is measured under a noisy environment, we have often witnessed that DeePC does not yield satisfactory performances, even with the regularization proposed in \citep{Coulson2019}.

As an alternative, we propose to employ the data-driven system representation by \cite{Persis2019}.
More specifically, the predicted output sequence for optimization at each sampling time is generated by the system equation, which is represented by the collected input/output data.
We will see in Section \ref{sec:sim} that the outcomes of this approach yield quite different output responses from those using DeePC under measurement noises.

However, employing the approach of \citep{Persis2019} to our purpose was not straightforward, and therefore, the contribution of this paper lies in the following points:
\begin{enumerate}
    
\item The approach of \citep{Persis2019} requires the knowledge of system order $n$. 
This may make sense when we can measure the system state in that the size of the state vector is the system order. 
However, since our interest is a completely model-free control, asking the knowledge of system order $n$ may be too much because it is a part of model information.
Our first contribution is to prove that the knowledge of $n$ is not necessary.
    
\item The way of handling multi-input-multi-output (MIMO) system in \citep{Persis2019} has a limitation (see Remark \ref{rem:mimo}).
Our second contribution is to present an idea of handling MIMO system as a multi-channel MISO (multi-input-single-output) system.
In this case, the multi-channel MISO system cannot be realized as the minimal order in general, but thanks to the derivation of the item (1) above, we can handle non-minimal order of system representation.
This idea enables the proposed method applicable to any MIMO systems without any limitation.

\item Our third contribution is a finding that the effect of measurement noise can be efficiently relieved by averaging the data-driven model of the system and by intentionally taking unnecessarily large $\bar n$ (the estimated upper bound of system order).
This effect will be observed in Section \ref{sec:sim}.
    
\end{enumerate}

{\bf Notation:} 
For a set of vectors $u_1, \dots, u_N$, we let $\text{col}(u_1, \dots ,u_N) := [u_1^T , \dots , u_N^T]^T$.
Given a discrete-time signal $u: \Z \to \R^m$, col$(u(a), \dots, u(b))$ is represented by $u_{[a, b]}$.\
The Kronecker product is written as $\otimes$.
The norm $\|u\|_R^2$ denotes the quadratic form $u^T R u$.
The $n \times n$ identity matrix is denoted by $I_n$ (or $I$ when no confusion is possible), and the $n \times m$ zero matrix is denoted by $0_{n \times m}$.
A vector $[0, 0, \dots, 0, 1, 0, \dots, 0]^T$ with the entry $1$ in the $i$-th place is denoted by $e_{i}$.
For a matrix $A$, $A^{\dagger}$ denotes the Moore-Penrose inverse of $A$.

\section{Problem Formulation and MPC}

We consider a discrete-time linear time-invariant system
\begin{align} \label{eq:sys0}
	\begin{split}
		x(t+1) &= A x(t) + B u(t), \quad
		y(t) = C x(t) 
	\end{split}
\end{align}
where $A \in \R^{n \times n}$, $B \in \R^{n \times m}$, $C \in \R^{p \times n}$, and $x(t) \in \R^n$, $u(t) \in \R^m$, $y(t) \in \R^p$ are the state, the control input, and the output at time $t$, respectively.
It is assumed that system \eqref{eq:sys0} is {\em controllable} and {\em observable}.
Given a reference $r(t) \in \R^p$, an input constraint set ${\mathcal U} \subset \R^m$, and an output constraint set ${\mathcal Y} \subset \R^p$, our goal is to build an output feedback controller such that $y(t)$ tracks $r(t)$ while satisfying the input and output constraints.
This goal should be achieved without the knowledge of system matrices $A$, $B$, $C$, and the system order $n$. 

\begin{Assumption}\label{ass:1}
The unknown system order $n$ belongs to a given interval $[1,\bar n]$ where $\bar n \in \Z$ is known.
\end{Assumption}

The assumption is easily met in many cases by taking sufficiently large $\bar n$.

When the system model \eqref{eq:sys0} is known, the goal is achievable straightforwardly by the model predictive control (MPC) with a state observer; that is, at each time $t$, get an estimate $\hat x(t)$ of the state, solve the optimization problem:
\begin{subequations}\label{eq:MPC}
\begin{align}
	\min_{\bar u} \ \ &\sum_{k=0}^{N-1} \left( \|\bar y_k - r(t+k)\|_Q^2 + \|\bar u_k \|_R^2 \right) \label{eq:a} \\
	\text{subject to}\ \ & \bar x_{k+1} = A \bar x_k + B \bar u_k, \ \bar y_k = C \bar x_k \label{eq:b} \\
	&\bar x_0 = \hat x(t) \\
	&\bar u_k \in {\mathcal U},\ \bar y_k \in {\mathcal Y}, \ k=0, 1, \dots, N-1
\end{align}
\end{subequations}
where $N \in \Z_{>0}$ is the time horizon, $\bar u = {\rm col}(\bar u_0,\dots,\bar u_{N-1})$, $Q \in \R^{p \times p}$ and $R \in \R^{m \times m}$ are positive semi-definite and positive definite matrices, respectively, and apply $u(t) = \bar u_0$ to the plant \eqref{eq:sys0} at time~$t$.

\begin{rem}
To ensure asymptotic convergence of $(y(t)-r(t))$ to zero, we need a reference input $u_r(t)$ that satisfies $x_r(t+1)=Ax_r(t)+Bu_r(t)$ and $r(t) = C x_r(t)$, $\forall t \ge 0$, with some trajectory $x_r(t)$, and the term $\|\bar u_k\|_R^2$ in the cost function in \eqref{eq:a} needs to be replaced with $\|\bar u_k - u_r(t+k)\|_R^2$.
However, computing $u_r$ is not always easy in the model-free setting, and thus, $u_r$ is often ignored in the literature.
\end{rem}

%%%%%%%%%%%%%%%%%%%%%%%%%
\section{Review of DeePC}

Data-enabled Predictive Control (DeePC) is firstly introduced in \citep{Coulson2019}, which is a neat and simple approach for model-free MPC.
Suppose that the system model \eqref{eq:sys0} is unknown but input/output data samples are available.
The following definition is a key to the forthcoming discussions.

%%%
\begin{defn}
	The signal $u_{[0,T-1]} \in \R^{mT}$ is {\em persistently exciting of order $L$} if the Hankel matrix
	\begin{equation*}
		{\mathcal H}_{L}(u):= \begin{bmatrix}
			u(0) & u(1) & \cdots & u(T-L) \\
			u(1) & u(2) & \cdots & \\
			\vdots & & \cdots & \\
			u(L-1) & u(L) & \cdots & u(T-1)
		\end{bmatrix}
	\end{equation*}
	has full row rank.
\end{defn}

Note that, if $u$ is persistently exciting of order $L$, then it is also persistently exciting of order $\tilde L$ for any $\tilde L \le L$.
Moreover, for a signal $u$ to be persistently exciting of order $L$, it is necessary that $T\ge (m+1)L-1$.

In order to introduce DeePC algorithm, let $T$, $T_{\rm ini}$, $N \in \Z_+$ be given such that $T \ge (m+1)(T_{\rm ini}+N+n)-1$.
We also let $u_d = \text{col}(u_d(0), \cdots, u_d(T-1))$ be a sequence of $T$ inputs applied to system \eqref{eq:sys0}, and $y_d = \text{col}(y_d(0), \cdots, y_d(T-1) )$ be the corresponding outputs. 
Here, the subscript $d$ indicates that it is the sample data collected offline from pre-experiments.
Define 
\begin{align*}
	\begin{bmatrix}
		U_p \\ U_f
	\end{bmatrix}:= {\mathcal H}_{T_{\rm ini}+N}(u_d),\
	\begin{bmatrix}
		Y_p \\ Y_f
	\end{bmatrix}:= {\mathcal H}_{T_{\rm ini}+N}(y_d)	
\end{align*}
where $U_p$ consists of the first $T_{\rm ini}$ block rows of ${\mathcal H}_{T_{\rm ini}+N}(u_d)$ and $U_f$ consists of the last $N$ block rows of ${\mathcal H}_{T_{\rm ini}+N}(u_d)$ ($Y_p$ and $Y_f$ are defined similarly).
Then, by the Fundamental Lemma \citep{Willems2005}, there exists a vector $g \in \R^{T-T_{\rm ini}-N+1}$ such that 
\begin{equation*}
	\begin{bmatrix}
		U_p \\ Y_p \\ U_f \\ Y_f \end{bmatrix} g =\begin{bmatrix}
		u_{[-T_{\rm ini}, -1]} \\ y_{[-T_{\rm ini}, -1]} \\ u_{[0, N-1]} \\ y_{[0, N-1]} 	
	\end{bmatrix},
\end{equation*}
if $u_d$ is persistently exciting of order $T_{\rm ini}+N+n$.
This implies that the system outputs $y_{[0, N-1]}$ can be computed provided that $u_d, y_d, u_{[-T_{\rm ini}, -1]}, y_{[-T_{\rm ini}, -1]}$ and $u_{[0, N-1]}$ are given.
In other words, future outputs $y_{[0, N-1]}$ can be predicted without the knowledge of system model \eqref{eq:sys0}.
Consider the optimization problem:
\begin{align} \label{eq:DeePC}
	\begin{split}
		\min_{g} \ \ &\sum_{k=0}^{N-1} \left( \| \bar y_{k}-r(t+k)\|_Q^2 + \| \bar u_{k} \|_R^2 \right) \\
		\text{subject to}\ \ & \begin{bmatrix}
			U_p \\ Y_p \\ U_f \\ Y_f \end{bmatrix} g =\begin{bmatrix}
			u_{\rm ini} \\ y_{\rm ini} \\ \bar u \\ \bar y 	\end{bmatrix} \\
		&u_{\rm ini}=\text{col}(u(t-T_{\rm ini}), \cdots, u(t-1)), \\
		&y_{\rm ini}=\text{col}(y(t-T_{\rm ini}), \cdots, y(t-1)),	\\
		&\bar u_k \in {\mathcal U},\ k=0, \cdots, N-1 \\
		&\bar y_k \in {\mathcal Y},\ k=0, \cdots, N-1,	
	\end{split}
\end{align}
where $\bar u = \text{col}(\bar u_0, \cdots, \bar u_{N-1})$ and $\bar y = \text{col}(\bar y_0, \cdots, \bar y_{N-1})$.
The DeePC algorithm is described as follows: at time $t-1$, the output $y(t-1)$ is measured, and construct $u_{\rm ini}$ and $y_{\rm ini}$ with $u(t-1)$. 
Solve the above optimization problem, get $\bar u = U_f g$ with the optimal solution $g$, and apply $u(t) = \bar u_0$ at time $t$.

To apply the DeePC algorithm to the case where the plant output is subject to measurement noise, the regularized DeePC (let us call it as rDeePC) is introduced based on the following optimization problem \citep{Coulson2019,Elokda2019,Berberich2020}: 
\begin{align}
	\min_{g,\sigma_y} \ \ \sum_{k=0}^{N-1} \big( & \| \bar y_{k}-r(t+k)\|_Q^2 + \| \bar u_{k} \|_R^2 \big)  + \lambda_g \| g\|^2 + \lambda_y \| \sigma_y \|^2 \notag \\
	\text{subject to}\ \ & \begin{bmatrix}
		U_p \\ Y_p \\ U_f \\ Y_f \end{bmatrix} g =\begin{bmatrix}
		u_{\rm ini} \\ y_{\rm ini} + \sigma_y \\ \bar u \\ \bar y 	\end{bmatrix}  \label{eq:rDeePC} \\
	&u_{\rm ini}=\text{col}(u(t-T_{\rm ini}), \cdots, u(t-1)), \notag \\
	&y_{\rm ini}=\text{col}(y(t-T_{\rm ini}), \cdots, y(t-1)),	\notag \\
	&\bar u_k \in {\mathcal U},\ k=0, \cdots, N-1, \notag \\
	&\bar y_k \in {\mathcal Y},\ k=0, \cdots, N-1,	\notag
\end{align}
where $\sigma_y \in \R^{p T_{\rm ini}}$ is a slack variable, and $\lambda_y, \lambda_g \in \R$ are regularization parameters.
Although the performance of rDeePC relies on the selection of $\lambda_g$ and $\lambda_y$, there is no systematic way to appropriately choose them.

\section{Data-Driven Predictive Control (D$^2$PC)}

Data-enabled Predictive Control (DeePC) is a simple approach for model-free MPC and it has been successfully applied to several practices. 
However, it often does not yield satisfactory performances when an unstable system is to be controlled or the output is measured under a noisy environment, even if the regularized DeePC \citep{Coulson2019,Elokda2019,Berberich2020} is employed.

As an alternative, one may build a system model \eqref{eq:sys0} from the experimental data, and plug the model in \eqref{eq:MPC}.
For the purpose of building a data-driven model, we employ the recent approach by \cite{Persis2019}.
Elimination of the use of state observer in the MPC \eqref{eq:MPC} is also from the idea of \citep[Section VI]{Persis2019}.
Let us call this strategy by Data-Driven Predictive Control (D$^2$PC).
However, this idea confronts an immediate difficulty that the plant order $n$ should be known in \citep{Persis2019}.
In this section, we briefly review the model building by \cite{Persis2019} and present how to overcome the difficulty.

\subsection{Data-driven representation of input and output: SISO case}

We first consider \eqref{eq:sys0} in the case of single-input-single-output (SISO); i.e., $m = p = 1$.
In this case, the input $u$ and the output $y$ of \eqref{eq:sys0} obeys
\begin{align} \label{eq:sys}
y(t) = - \sum_{j=1}^n a_j y(t-j) + \sum_{j=1}^n b_j u(t-j)
\end{align}
where the coefficients satisfy
\begin{gather}\label{eq:twopoly}
\begin{split}
z^n + a_1 z^{n-1} + \dots + a_n = \det(zI - A) \\
b_1 z^{n-1} + \dots + b_n = \det\left( \begin{bmatrix} zI - A &\;\; -B \\ C & 0 \end{bmatrix} \right) .
\end{split}
\end{gather}
Define a vector $\chi \in \R^{2n}$ as
\begin{align*}% \label{eq:chi}
\chi(t) := \text{col}(&y(t-n), y(t-n+1), \dots, y(t-1), \\
&u(t-n), u(t-n+1), \dots, u(t-1))
\end{align*}
which is available to the controller for all $t \ge n$ because $y$ is measured and $u$ is generated by the controller.
Then, it is seen that 
\begin{align}\label{eq:ABsiso}
\begin{split}
\chi(t+1) &= {\mathcal A} \chi(t) + {\mathcal B} u(t), \quad
y(t) = {\mathcal C} \chi(t)
\end{split}
\end{align}
where $\AAA \in \R^{2n \times 2n}$, $\BB \in \R^{2n \times 1}$, and ${\mathcal C} \in \R^{1 \times 2n}$ are 
\begin{align*}
		{\mathcal A} & = \begin{bmatrix}
			0 & 1 & \dots & 0 & 0 & 0  & \dots  & 0 \\
			0 & 0 & \dots & 0 & 0 & 0  & \dots  & 0 \\
			\vdots & \vdots &  \vdots & \vdots & \vdots  &  & \vdots\\
			0 & 0 & \dots & 1 & 0 & 0  & \dots  & 0 \\
			-a_n & -a_{n-1} & \dots  & -a_1 & b_{n} & b_{n-1} &\dots & b_1\\
			0 & 0 & \dots & 0 & 0 & 1 & \dots  & 0 \\
			0 & 0 & \dots & 0 & 0 & 0  & \dots  & 0 \\
			\vdots & \vdots &  & \vdots & \vdots & \vdots   & \vdots\\
			0 & 0 & \dots & 0 & 0 & 0  & \dots  & 1 \\			
			0 & 0 & \dots & 0 & 0 & 0  & \dots  & 0 
		\end{bmatrix} \\
		{\mathcal B} & = [0 \ \ 0 \ \dots \ 0 \ \ 0 \ \ 0 \ \ 0 \ \dots \ 0 \ \ 1]^T \\
{\mathcal C} &= [-a_n \ \ -a_{n-1} \ \dots \ -a_1 \ \ b_{n} \ \ b_{n-1} \ \dots \ \ b_1 ] .
\end{align*}
It is noted that both \eqref{eq:sys0} and \eqref{eq:ABsiso} yield the same inputs and outputs for corresponding initial conditions.
Note also that, while the system \eqref{eq:sys0} is controllable and observable, the pair $(\AAA,{\mathcal C})$ is not observable, which means that \eqref{eq:ABsiso} is a non-minimal realization of \eqref{eq:sys0}.
Nevertheless, system \eqref{eq:ABsiso} can remain controllable as follows, whose proof is found in \citep[Lemma 3.4.7]{Goodwin2014}. 

\begin{lem}[\cite{Persis2019}] \label{lem:ctrb}
With two polynomials in \eqref{eq:twopoly} being coprime, the pair $(\AAA, \BB)$ is controllable.
\end{lem}

Now suppose that we know $n$, and from this, suppose that an experiment is performed and the data $u_d(t)$ and $y_d(t)$ are collected for $T+n$ steps, where $T \ge 4n+1$.
Here, we appended subscript $d$ to indicate they are input/output data from a pre-experiment before the actual run of the control.
From the data, one can obtain
\begin{align}\label{eq:XXU}
\begin{split}
\XX_- &:= [\chi_d(0) \ \chi_d(1) \ \dots \ \chi_d(T-1)], \\
\XX_+ &:= [\chi_d(1) \ \chi_d(2) \ \dots \ \chi_d(T)], \\
U_- &:= [u_d(0) \ u_d(1) \ \dots \ u_d(T-1)]
\end{split}
\end{align}
where $\chi_d(t) = \text{col}(y_d(t-n),\dots,y_d(t-1),u_d(t-n),\dots,u_d(t-1)) \in \R^{2n}$.
The input $u_d$ is assumed to be persistently exciting of order $2n+1$, which implies that, with the controllability of $(\AAA,\BB)$,
the data matrix
\begin{equation}\label{eq:fullrowrank}
\begin{bmatrix} \XX_- \\ U_- \end{bmatrix} \in \R^{(2n+1) \times T} \quad \text{has full row rank}
\end{equation}
(see \citep[Corollary 2]{Willems2005} for a proof).
This is the key to the identification presented in \citep{Persis2019} because
\begin{equation}\label{eq:forall}
\text{$\forall \chi(t), u(t)$, $\exists g(t) \in \R^{T}$ such that}
\; \begin{bmatrix} \chi(t) \\ u(t) \end{bmatrix} = \begin{bmatrix} \XX_- \\ U_- \end{bmatrix} g(t).
\end{equation}
This means that 
\begin{equation}\label{eq:gt}
g(t) = \underbrace{\begin{bmatrix} \XX_- \\ U_- \end{bmatrix}^\dagger \begin{bmatrix} \chi(t) \\ u(t) \end{bmatrix}}_{=:g_1(t)} + \underbrace{\left(I - \begin{bmatrix} \XX_- \\ U_- \end{bmatrix}^\dagger \begin{bmatrix} \XX_- \\ U_- \end{bmatrix} \right) w}_{=:g_2(t)}
\end{equation}
for arbitrary $w \in \R^T$, where the second term spans the null space of $\text{col}(\XX_-,U_-)$.
Then, we have, for any $w \in \R^T$,
\begin{equation*}
\begin{bmatrix} \XX_- \\ U_- \end{bmatrix} g_2(t) = \left(\begin{bmatrix} \XX_- \\ U_- \end{bmatrix} - \begin{bmatrix} \XX_- \\ U_- \end{bmatrix} \begin{bmatrix} \XX_- \\ U_- \end{bmatrix}^\dagger \begin{bmatrix} \XX_- \\ U_- \end{bmatrix} \right) w = 0
\end{equation*}
and, by \eqref{eq:ABsiso},
\begin{equation}\label{eq:ABequation}
\XX_+ = [\AAA, \BB] \begin{bmatrix} \XX_- \\ U_- \end{bmatrix}. \quad 
\end{equation}
Thus, it follows that 
\begin{align}
	\chi(t+1) &= [\AAA, \BB] \begin{bmatrix} \chi(t) \\ u(t) \end{bmatrix} = [\AAA, \BB] \begin{bmatrix} \XX_- \\ U_- \end{bmatrix} g(t) \notag \\
	&= \XX_+ \begin{bmatrix} \XX_- \\ U_- \end{bmatrix}^\dagger \begin{bmatrix} \chi(t) \\ u(t) \end{bmatrix} \label{eq:dbrep} \\
	y(t) &= e_n^\top \chi(t+1) = e_n^\top \XX_+ \begin{bmatrix} \XX_- \\ U_- \end{bmatrix}^\dagger \begin{bmatrix} \chi(t) \\ u(t) \end{bmatrix} . \notag
\end{align}
Therefore, identification of \eqref{eq:ABsiso} is done, and \eqref{eq:dbrep} is a data-driven representation of \eqref{eq:ABsiso}.
Then, \eqref{eq:b} is replaced with \eqref{eq:dbrep}, and the MPC \eqref{eq:MPC} can be employed with the role of $\hat x$ being played by $\chi$.

Unfortunately, the discussion so far is based on knowledge of the plant's order $n$.
Now, our treatment begins with the observation that, with $\bar n \ge n$, the input/output of the plant \eqref{eq:sys0} still satisfies
\begin{align} \label{eq:inandout2}
y(t) = - \sum_{j=1}^{\bar n} \bar a_j y(t-j) + \sum_{j=1}^{\bar n} \bar b_j u(t-j).
\end{align}
In this case, however, the coefficients $\bar a_j$ and $\bar b_j$ are not unique, and two polynomials
\begin{equation}\label{eq:twopoly3}
z^{\bar n} + \bar a_1 z^{\bar n-1} + \dots + \bar a_{\bar n} \quad \text{and} \quad 
\bar b_1 z^{\bar n-1} + \dots + \bar b_{\bar n} .
\end{equation}
are never coprime, and the common roots of two polynomials correspond to cancelled poles and zeros when the transfer function is constructed.
Proceeding similarly as before, define a vector $\bar \chi \in \R^{2\bar n}$ as
\begin{align} \label{eq:chi2}
\begin{split}
\bar \chi(t) := \text{col} (&y(t-\bar n), y(t-\bar n+1), \dots, y(t-1), \\
&u(t-\bar n), u(t-\bar n+1), \dots, u(t-1) ).
\end{split}
\end{align}
Then, we have 
\begin{align}\label{eq:ABsisobar}
\begin{split}
\bar \chi(t+1) &= \bar \AAA \bar \chi(t) + \bar \BB u(t), \qquad % \\
y(t) = \bar {\mathcal C} \bar \chi(t)
\end{split}
\end{align}
where $\bar \AAA \in \R^{2\bar n \times 2\bar n}$, $\bar \BB \in \R^{2\bar n \times 1}$, and $\bar {\mathcal C} \in \R^{1 \times 2 \bar n}$ have the same structure as $\AAA$, $\BB$, and ${\mathcal C}$ in \eqref{eq:ABsiso}, with $a_j$ and $b_j$ replaced by $\bar a_j$ and $\bar b_j$, respectively.
Also let $\bar \XX_-$ and $\bar \XX_+$ be defined similarly as \eqref{eq:XXU} with $\chi_d \in \R^{2n}$ replaced by $\bar \chi_d \in \R^{2\bar n}$, and $T \ge 4\bar n + 1$.
For instance, $\bar {\mathcal{X}}_-$ is given by
	\begin{align*}
		\bar \XX_- = \begin{bmatrix}
			y_d(-\bar n) & y_d(-\bar n+1) & \dots & y_d(-\bar n+T-1) \\
			y_d(-\bar n+1) & y_d(-\bar n+2) & \dots & y_d(-\bar n+T) \\
			\vdots & \vdots & \ddots & \vdots \\
			y_d(-1) & y_d(0) & \dots & y_d(T-2) \\
			u_d(-\bar n) & u_d(-\bar n+1) & \dots & u_d(-\bar n+T-1) \\
			u_d(-\bar n+1) & u_d(-\bar n+2) & \dots & u_d(-\bar n+T) \\
			\vdots & \vdots & \ddots & \vdots \\
			u_d(-1) & u_d(0) & \dots & u_d(T-2) 						
		\end{bmatrix}.
	\end{align*}

However, loss of coprimeness in \eqref{eq:twopoly3} incurs loss of controllability for the pair $(\bar \AAA,\bar \BB)$,
and loss of controllability means that the matrix 
\begin{equation}\label{eq:fullrowrank2}
\begin{bmatrix} \bar \XX_- \\ U_- \end{bmatrix} \in \R^{(2\bar n+1) \times T}
\end{equation}
no longer has
%does not have 
full row rank, even if $u_d$ is sufficiently rich (i.e., persistently exciting of arbitrary order).
Nevertheless, we claim that 
\eqref{eq:forall} still holds {\em for the pair $\bar \chi(t)$ and $u(t)$ satisfying \eqref{eq:ABsisobar}},
which is the first contribution of this paper.
More specifically, we have the following.

\begin{lem}\label{lem:key1}
Under Assumption \ref{ass:1}, suppose that $u_d$ is persistently exciting of order $2 \bar n + 1$.
If $\bar \chi$ and $u$ satisfy \eqref{eq:ABsisobar}, then, for each $t \ge 0$, there is $\bar g(t) \in \R^{T}$ such that
\begin{equation}\label{eq:forall2}
\begin{bmatrix} \bar \chi(t) \\ u(t) \end{bmatrix} = \begin{bmatrix} \bar \XX_- \\ U_- \end{bmatrix} \bar g(t).
\end{equation}
\end{lem}

{\it Proof:} 
Let us define an intermediate variable $\hat \chi(t) \in \R^{n+\bar n}$ as $\chi(t-(\bar n-n))$ with more input samples appended; that is, 
\begin{align*}
\hat \chi(t) = \text{col}(&\chi(t-\bar n + n), \\ 
&u(t-\bar n + n), u(t- \bar n + n + 1), \dots, u(t-1) ) \\
= \text{col}(&y(t-\bar n), y(t-\bar n+1), \dots, y(t-\bar n+n-1), \\ 
&u(t-\bar n), u(t-\bar n+1), \dots, u(t-1) ) .
\end{align*}
Then, it follows that
\begin{align*}
&\hat \chi (t+1) = \begin{bmatrix}
\AAA \chi(t-\bar n +n) + \BB u(t-\bar n +n) \\ u(t-\bar n +n+1) \\ u(t- \bar n +n +2) \\ \vdots \\ u(t) \end{bmatrix} \\
&= \begin{bmatrix} \AAA & \hat \AAA_{1,2} \\ 0_{(\bar n -n) \times 2n } & \hat \AAA_{2,2} \end{bmatrix}
\begin{bmatrix} \chi(t-\bar n +n) \\ u(t-\bar n +n) \\ u(t- \bar n +n +1) \\ \vdots \\ u(t-1) \end{bmatrix} 
+ \begin{bmatrix} 0_{2n \times 1} \\ 0 \\ \vdots \\ 0 \\ 1 \end{bmatrix} u(t) \\
&=: \hat \AAA \hat \chi(t) + \hat \BB u(t) 
\end{align*}
where $\hat \AAA_{1,2} = [\BB \ 0_{2n \times (\bar n-n-1)}]$ and the $(i,j)$-th component of $\hat \AAA_{2,2}$ is 1 if $j = i+1$, and 0 otherwise.

Then, (even if $(\bar\AAA,\bar\BB)$ is not controllable) it is seen that $(\hat \AAA,\hat \BB)$ is controllable by the PBH rank test.
Indeed, the matrix 
\begin{align*}
&\begin{bmatrix}
sI - \hat \AAA & \hat \BB
\end{bmatrix} = \begin{bmatrix}
		sI - \AAA & -\BB & 0_{2n \times 1} & 0_{2n \times 1} & \dots & 0_{2n \times 1} & 0_{2n \times 1} \\
		0_{1 \times 2n} & s & -1 & 0 & \dots & 0 & 0\\
		0_{1 \times 2n} & 0 & s  & -1 & \dots & 0 & 0\\
		\dots \\
		0_{1 \times 2n} & 0 & 0  & 0 & \dots & s & 1\\		
	\end{bmatrix}
\end{align*}
has full row rank for all $s \in \mathbb{C}$, because $[sI - \AAA, \; -\BB]$ has full row rank for all $s \in \mathbb{C}$.
Therefore, with $\hat \XX_- := [\hat \chi_d(0) \ \hat \chi_d(1) \ \dots \ \hat \chi_d(T-1)]$,
it follows from \citep{Willems2005} that
\begin{equation*} 
\begin{bmatrix} \hat \XX_- \\ U_- \end{bmatrix} \quad \text{has full row rank}
\end{equation*} 
because $u_d$ is persistently exciting of order $\bar n + n+1$.
This in turn implies that there exists a vector $\bar g(t) \in \R^T$ such that
\begin{align*}
\begin{bmatrix}
y(t-\bar n) \\ \vdots \\ y(t-\bar n+n-1) \\ u(t-\bar n) \\ \vdots \\ u(t)
\end{bmatrix}
= \begin{bmatrix} \hat \XX_- \\ U_- \end{bmatrix} \bar g(t).
\end{align*}
By the definition of $\hat \XX_-$ and $U_-$, it follows that 
\begin{align}\label{eq:tmp0220}
\begin{bmatrix} y(t-\bar n) \\ \vdots \\ y(t-\bar n+n-1) \\ u(t-\bar n) \\ \vdots \\ u(t) \end{bmatrix} 
= \sum_{i=1}^{T} \begin{bmatrix}
y_d(-\bar n-1+i) \\ \vdots \\ y_d(-\bar n + n-2 +i) \\ u_d (-\bar n-1+i) \\ \vdots \\ u_d(-1+i)	\end{bmatrix} \bar g_i(t)
\end{align}
where $\bar g_i$ is the $i$-th component of $\bar g$.
From \eqref{eq:sys} and \eqref{eq:tmp0220}, 
\begin{align*}
y(t &- \bar n + n) \\
&= \sum_{j=1}^{n} -a_{j} y(t-\bar n+n-j) +  \sum_{j=1}^{n} b_{j} u(t-\bar n+n-j) \\
&= \sum_{j=1}^{n} -a_{j} \left( \sum_{i=1}^T y_d(-\bar n+n-j+i-1) \bar g_i(t) \right) \\
&\quad + \sum_{j=1}^{n} b_{j} \left( \sum_{i=1}^T u_d(-\bar n+n-j+i-1) \bar g_i(t) \right) \\
&= \sum_{i=1}^T \left( \sum_{j=1}^{n} -a_{j} y_d(-\bar n+n-j+i-1)\right) \bar g_i(t) \\
&\quad + \sum_{i=1}^T \left( \sum_{j=1}^{n} b_{j} u_d(-\bar n+n-j+i-1)\right) \bar g_i(t) \\
&= \sum_{i=1}^T y_d (-\bar n+n-1+i) \bar g_i(t) ,
\end{align*}
which can be appended to \eqref{eq:tmp0220} yielding
\begin{align}\label{eq:tmp0421}
\begin{bmatrix}
y(t-\bar n) \\ \vdots \\ y(t-\bar n+n-1) \\ y(t-\bar n+n) \\ u(t-\bar n) \\ \vdots \\ u(t)
\end{bmatrix} = \sum_{i=1}^{T} \begin{bmatrix}
y_d(-\bar n-1+i) \\ \vdots \\ y_d(-\bar n + n-2 +i) \\ y_d(-\bar n + n-1 +i) \\ u_d (-\bar n -1+i) \\ \vdots \\ u_d(-1+i)	
\end{bmatrix} \bar g_i(t).
\end{align}
Similarly, by \eqref{eq:sys} and \eqref{eq:tmp0421}, it is seen that $y(t-\bar n+n+1) = \sum_{i=1}^T y_d (-\bar n+n+i) \bar g_i(t)$ and that 
\begin{align*}
	\begin{bmatrix}
		y(t-\bar n) \\ \vdots \\ y(t-\bar n+n) \\ y(t-\bar n+n+1) \\ u(t-\bar n) \\ \vdots \\ u(t) \end{bmatrix} 
	= \sum_{i=1}^{T} \begin{bmatrix}
		y_d(-\bar n-1+i) \\ \vdots \\ y_d(-\bar n + n-1 +i) \\ y_d(-\bar n + n +i) \\ u_d (-\bar n -1+i) \\  \vdots \\ u_d(-1+i) \end{bmatrix} \bar g_i(t).
\end{align*}
Repeating this procedure $\bar n - n -2$ times more, the left-hand side and the matrix in the right-hand side grow to $\text{col}(\bar \chi(t), u(t))$ and $\text{col}(\bar \XX_-, U_-)$, respectively, yielding \eqref{eq:forall2}.
\hfill $\blacksquare$

\begin{theorem}\label{thm:siso}
Consider a single-input-single-output system in the form \eqref{eq:sys0}.
Under Assumption \ref{ass:1}, suppose that $u_d$ is persistently exciting of order $2\bar n+1$.
Then, the input $u(t)$ and the output $y(t)$ of \eqref{eq:sys0} are governed by a data-driven representation:
\begin{align} \label{eq:main}
\begin{split}
\bar \chi(t+1) &= \bar \XX_+ \begin{bmatrix} \bar \XX_- \\ U_- \end{bmatrix}^{\dagger} 
\begin{bmatrix} \bar \chi(t) \\ u(t) \end{bmatrix} \\
y(t) &= e_{\bar n}^\top \bar \chi(t+1) = e_{\bar n}^\top \bar \XX_+ \begin{bmatrix} \bar \XX_- \\ U_- \end{bmatrix}^{\dagger} 
\begin{bmatrix} \bar \chi(t) \\ u(t) \end{bmatrix}
\end{split}
\end{align}
where $\bar \chi \in \R^{2\bar n}$.
\end{theorem}

We note that \eqref{eq:main} need not be the same as \eqref{eq:ABsisobar}.
This is clear because $\bar\AAA$, $\bar\BB$, and $\bar{\mathcal C}$ in \eqref{eq:ABsisobar} are not unique.
Therefore, \eqref{eq:main} is simply one of the suitable representations between inputs and outputs, and it is not an identification of \eqref{eq:ABsisobar}.

{\it Proof:}
For notational simplicity, let 
\begin{equation}
J:=\begin{bmatrix} \bar \XX_- \\ U_- \end{bmatrix} \quad \text{and} \quad {v(t)}:= \begin{bmatrix}
\bar \chi(t) \\ u(t)  
\end{bmatrix}.
\end{equation}
Then, since  $J = J J^{\dagger} J$ and $\bar g(t)$ is a particular solution to \eqref{eq:forall2}, it follows that 
$$v(t) = J \bar g(t) = J J^{\dagger} J \bar g(t) =J J^{\dagger} v(t).$$
As a result, 
a general solution to \eqref{eq:forall2} is of the form 
\begin{align}
	J^{\dagger} {v(t)} + (I-J^{\dagger}J)w(t)
\end{align}
where $w(t) \in \R^T$ is an arbitrary vector.
Thus, there exists $\bar w(t) \in \R^T$ such that 
\begin{align}
	\bar g(t) = J^{\dagger} {v(t)} + (I-J^{\dagger}J) \bar w(t),
\end{align}
which, together with \eqref{eq:ABsisobar}, implies  
\begin{align}\label{eq:temp}
\begin{split}
	\bar \chi(t+1) &= [\bar \AAA \ \ \bar \BB ] \begin{bmatrix}
		\bar \chi (t) \\ u(t)
	\end{bmatrix}  
	=[\bar \AAA \ \ \bar \BB ]  J \bar g(t) \\
	&=[\bar \AAA \ \ \bar \BB ]  J J^{\dagger} {v(t)} + [\bar \AAA \ \ \bar \BB ]  J(I-J^{\dagger}J)\bar w(t) \\
	&=\bar \XX_+ J^{\dagger} {v(t)}
\end{split}
\end{align}
where the last equality follows from the fact that $[\bar \AAA \ \ \bar \BB ]  J = \bar \XX_+$ and $J = J J^{\dagger} J$.
Therefore, \eqref{eq:main} is established.
\hfill $\blacksquare$

\subsection{Data-driven representation: MIMO case}

Our treatment of multi-input-multi-output (MIMO) case is to split the output channels and to handle the MIMO system \eqref{eq:sys0} as $p$ parallel multi-input-single-ouput (MISO) systems, which is the second contribution of this paper.
That is, from
$$y(z) = \begin{bmatrix} G_1(z) \\ \vdots \\ G_p(z) \end{bmatrix} u(z) = C(zI-A)^{-1}B u(z)$$
where $G_i(z)$ is $1$-by-$m$ transfer function matrix whose elements are coprime transfer functions, another realization of \eqref{eq:sys0} is
\begin{align}\label{eq:nonmin}
\begin{split}
\tilde x(t+1) &= \begin{bmatrix} \tilde A_1 & \dots & 0 \\ \vdots & \ddots & \vdots \\ 0 & \dots & \tilde A_p \end{bmatrix} \tilde x(t) + \begin{bmatrix} \tilde B_1 \\ \vdots \\ \tilde B_p \end{bmatrix} u(t) \\
y(t) &= \begin{bmatrix} \tilde C_1 & \dots & 0 \\ \vdots & \ddots & \vdots \\ 0 & \dots & \tilde C_p \end{bmatrix} \tilde x(t)
\end{split}
\end{align}
where $\tilde A_i \in \R^{\tilde n_i \times \tilde n_i}$, $\tilde B_i \in \R^{\tilde n_i \times m}$, and $\tilde C_i \in \R^{1 \times \tilde n_i}$, and $\tilde n_i$ is the order of the least common multiple of the denominator polynomials of $G_i(z)$.
Therefore, each $(\tilde A_i,\tilde B_i,\tilde C_i)$ is a {\em minimal} realization of $G_i(z)$, but $\sum_{i=1}^p \tilde n_i \ge n$ in general so that \eqref{eq:nonmin} is possibly a non-minimal realization of \eqref{eq:sys0}.

\begin{Assumption}\label{ass:1prime}
The unknowns $\tilde n_i$, $i=1,\dots,p$, belong to a given interval $[1,\bar n]$ where $\bar n \in \Z$ is known.
\end{Assumption}

From the discussions so far, we know that there are $a_{i,j} \in \R$ and $b_{i,j} \in \R^{1 \times m}$ such that 
\begin{align}\label{eq:MIsys}
y_i(t) = - \sum_{j=1}^{\tilde n_i} a_{i,j} y_i(t-j) + \sum_{j=1}^{\tilde n_i} b_{i,j} u(t-j), \quad i=1,\dots,p
\end{align}
which corresponds to the relation $y_i(z) = G_i(z) u(z)$.
Treating this relation as \eqref{eq:sys}, the following (non-minimal) relation (corresponding to \eqref{eq:inandout2}) also holds true:
\begin{align} \label{eq:inandout3}
y_i(t) = - \sum_{j=1}^{\bar n} \bar a_{i,j} y_i(t-j) + \sum_{j=1}^{\bar n} \bar b_{i,j} u(t-j)
\end{align}
where $\bar a_{i,j} \in \R$ and $\bar b_{i,j} \in \R^{1 \times m}$.
The rest of the development proceeds similarly to the SISO case.
In particular, we have the following result.

\begin{theorem}\label{thm:miso}
Consider the system \eqref{eq:sys0}.
Under Assumption \ref{ass:1prime}, suppose that $u_d \in \R^m$ is persistently exciting of order $2 \bar n+1$.
Then, the input $u(t) \in \R^m$ and the output $y(t) \in \R^p$ are governed by a data-driven representation:
\begin{align} \label{eq:main2}
\begin{split}
&\bar \chi_i(t+1) = \bar \XX_{i,+} \begin{bmatrix} \bar \XX_{i,-} \\ U_- \end{bmatrix}^{\dagger} 
\begin{bmatrix} \bar \chi_i(t) \\ u(t) \end{bmatrix} \\
&y_i(t) = e_{\bar n}^\top \bar \chi_i(t+1) = e_{\bar n}^\top \bar \XX_{i,+} \begin{bmatrix} \bar \XX_{i,-} \\ U_- \end{bmatrix}^{\dagger} 
\begin{bmatrix} \bar \chi_i(t) \\ u(t) \end{bmatrix}
\end{split}
\end{align}
for $i=1,\dots,p$, where 
\begin{align*}
\bar \chi_i(t) &:= {\rm col} (y_i(t-\bar n), y_i(t-\bar n+1), \dots, y_i(t-1), \\
&\quad u(t-\bar n), u(t-\bar n+1), \dots, u(t-1) ) \quad \in \R^{(1+m)\bar n} \\
\bar \XX_{i,-} &:= [\bar\chi_{d,i}(0) \ \bar\chi_{d,i}(1) \ \dots \ \bar\chi_{d,i}(T-1)] \\
\bar \XX_{i,+} &:= [\bar\chi_{d,i}(1) \ \bar\chi_{d,i}(2) \ \dots \ \bar\chi_{d,i}(T)] \\
U_- &:= [u_d(0) \ u_d(1) \ \dots \ u_d(T-1)]
\end{align*}
in which the subscript $d$ implies they are the data obtained from a pre-experiment, and $T \ge 4\bar n+1$.
\end{theorem}

{\it Proof:}
Let $b_{i,j,k} \in \R$ be the $k$-th component of the row vector $b_{i,j}$, and let
\begin{align*} 
\chi_i(t) &:= \text{col} ( y_i(t-\tilde n_i), y_i(t-\tilde n_i+1), \dots, y_i(t-1), \\
&\hspace{11mm} u(t-\tilde n_i), u(t-\tilde n_i+1), \dots, u(t-1) ).
\end{align*}
Then, it is easy to see that 
\begin{equation}
\chi_i(t+1) = \AAA_i \chi_i(t) + \BB_i u(t)
\end{equation}
where $\AAA_i \in \R^{(m+1)\tilde n_i \times (m+1)\tilde n_i}$ and $\BB_i \in \R^{(m+1)\tilde n_i \times m}$ are defined similarly as in \eqref{eq:ABsiso}.	
Since each elements of $G_i(z)$ are coprime transfer functions, 
$A_i(z) := z^{\tilde n_i} + a_{i,1}z^{\tilde n_i-1} + \dots + a_{i,\tilde n_i}$ and $B_{i,k}(z) := b_{i,1,k} z^{\tilde n_i-1} + \dots + b_{i,\tilde n_i,k}$, $1 \le k \le m$, do not have a common divisor.
As a result, it can be shown by using \citep[Lemma 3.4.7]{Goodwin2014} that $(\AAA_i, \BB_i)$ is controllable.
Let
\begin{multline} \label{MIchi3}
\hat \chi_i(t) := \text{col} ( y_i(t-\bar n), y_i(t-\bar n+1), \dots, y(t-\bar n + \tilde n_i - 1), \\
u(t-\bar n), u(t-\bar n+1), \dots, u(t-1) ).
\end{multline}
Then, since 
\begin{align*}
\hat \chi_i(t) &= \text{col}( \chi_i(t-\bar n+\tilde n_i), \\
&\qquad u(t-\bar n+\tilde n_i), u(t-\bar n+\tilde n_i+1), \dots, u(t-1) )
\end{align*}
it is seen that
\begin{align}\label{MIchi_hat}
\hat \chi_i(t+1) & = \hat \AAA_i \hat \chi_i(t) + \hat \BB_i u(t),
\end{align}	
where, with $\Delta_i := \bar n-\tilde n_i$,
\begin{align*}
\hat \AAA_i &= \begin{bmatrix}
	\AAA_i & \BB_i & 0_{(m+1)\tilde n_i \times (\Delta_i-1)m}  \\
	0_{(\Delta_i-1) m \times (m+1)\tilde n_i}
	& 0_{(\Delta_i-1) m \times m}  & I_{(\Delta_i-1)m} \\
	0_{m \times (m+1)\tilde n_i}
	& 0_{m \times m}  & 0_{m \times (\Delta_i-1)m}
	\end{bmatrix}, \\ 
	\hat \BB_i &= \begin{bmatrix} 
	0_{(m+1)\tilde n_i \times m}  \\ 0_{(\Delta_i-1)m \times m} \\
	I_m 
	\end{bmatrix}.
\end{align*} 
Since $(\AAA_i, \BB_i)$ is controllable, $(\hat \AAA_i, \hat \BB_i)$ is also controllable. 

Now, for any $\bar \chi_i(t)$ and $u(t)$ constructed from \eqref{eq:inandout3}, we claim that there exists $\bar g(t) \in \R^T$, for each $t \ge 0$, such that 
\begin{equation} \label{eq:Ax_b2}
\begin{bmatrix}	\bar \chi_i (t) \\ u(t) \end{bmatrix} = \begin{bmatrix}	\bar \XX_{i,-} \\ U_-   \end{bmatrix} \bar g(t).
\end{equation} 
Since $(\hat \AAA_i, \hat \BB_i)$ is controllable and $u_d$ is persistently exciting of order $\tilde n_i + \bar n + 1$, it follows from \citep{Willems2005} that 
\begin{equation*} 
\begin{bmatrix}	\hat \XX_{i,-} \\ U_- \end{bmatrix} \ \text{has full row rank,}
\end{equation*}
where $\hat \XX_{i,-}$ is defined similarly as in $\bar \XX_{i,-}$.
Thus, there exists a vector $\bar g = \text{col} (\bar g_1, \dots , \bar g_T) \in \R^{T}$ such that 
\begin{align}\label{MIux_tmp1}
\begin{split}
&\begin{bmatrix}
y_i(t-\bar n) \\ \vdots \\ y_i(t-\bar n+\tilde n_i-1) \\ u_1(t-\bar n) \\ u_2(t-\bar n) \\ \vdots \\ u_{m-1}(t) \\ u_m(t) \end{bmatrix}
= \sum_{l=1}^{T} \begin{bmatrix}
y_{d,i}(-\bar n-1+l) \\ \vdots \\ y_{d,i}(-\bar n + \tilde n_i - 2 + l) \\ 
u_{d,1} (-\bar n -1+l) \\ u_{d,2} (-\bar n -1+l) \\ \vdots \\ u_{d,m-1}(l-1) \\ u_{d,m}(l-1) 	
\end{bmatrix} \bar g_l .  
\end{split}
\end{align}			
From \eqref{eq:MIsys} and \eqref{MIux_tmp1}, it can be shown that 
\begin{align*}
y_i(t-\bar n+\tilde n_i) &= \sum_{l=1}^T  y_{d,i} (-\bar n+\tilde n_i-1+l) \bar g_l,
\end{align*}		
which implies that 
\begin{align*}
&\begin{bmatrix}
y_i(t-\bar n) \\ \vdots \\ y_i(t-\bar n+\tilde n_i) \\ u_1(t-\bar n) \\ u_2(t-\bar n) \\ \vdots \\ u_{m-1}(t) \\ u_m(t) \end{bmatrix}
= \sum_{l=1}^{T} \begin{bmatrix}
y_{d,i}(-\bar n-1+l) \\ \vdots \\ y_{d,i}(-\bar n + \tilde n_i - 1 + l) \\ 
u_{d,1} (-\bar n -1+l) \\ u_{d,2} (-\bar n -1+l) \\ \vdots \\ u_{d,m-1}(l-1) \\ u_{d,m}(l-1) 	
\end{bmatrix} \bar g_l .  
\end{align*}
Continuing in this way, \eqref{eq:Ax_b2} can be established.
Finally, since $\bar g$ is a particular solution to \eqref{eq:Ax_b2}, the theorem is proved by the same method as in Theorem \ref{thm:siso}.  
\hfill
$\blacksquare$

\begin{rem}\label{rem:mimo}
The idea of handling a MIMO system as $p$ parallel MISO systems is useful even when the system order $n$ is known.
As a matter of fact, a way to handle MIMO systems was presented in \citep[Section VI.C]{Persis2019}, which is however not applicable in some cases, while the proposed method is always applicable.
To appreciate this point, let us consider an example system \eqref{eq:sys0} with
\begin{align*}
A = \begin{bmatrix} 0 & 1 \\ 0 & 0 \end{bmatrix}, \quad 
B = \begin{bmatrix} 0  \\ 1 \end{bmatrix}, \quad
C = \begin{bmatrix}	1 & 0 \\ 1 & 1 \end{bmatrix}
\end{align*}	
which is controllable and observable.
Following the treatment in \citep[Section VI.C]{Persis2019}, one finds $A_i \in \R^{2 \times 2}$ and $B_i \in \R^{2 \times 1}$ such that
$$y(t) = -A_1y(t-1)-A_2y(t-2)+B_1u(t-1)+B_2u(t-2).$$
Then, with the knowledge of $n=2$, define $\chi(t) = {\rm col}(y(t-2),y(t-1),u(t-2),u(t-1)) \in \R^{(m+p)n}$ where $(m+p)n = 6$.
Then, $\chi(t)$ satisfies the relation \eqref{eq:ABsiso} with
\begin{align*}
\AAA &= \begin{bmatrix} 0_{2 \times 2} & I_2 & 0_{2 \times 1} & 0_{2 \times 1} \\ -A_2 & -A_1 & B_2 & B_1 \\ 0_{1 \times 2} & 0_{1 \times 2} & 0 & 1 \\ 0_{1 \times 2} & 0_{1 \times 2} & 0 & 0 \end{bmatrix} = \begin{bmatrix} 0 & 0 & 1 & 0 & 0 & 0 \\
0 & 0 & 0 & 1 & 0 & 0 \\
0 & 0 & 0 & 0 & 1 & 0 \\
0 & 0 & 0 & 0 & 1 & 1 \\
0 & 0 & 0 & 0 & 0 & 1 \\
0 & 0 & 0 & 0 & 0 & 0 \end{bmatrix}, \\
\BB &= \begin{bmatrix} 0 & 0 & 0 & 0 & 0 & 1 \end{bmatrix}^T, \\
{\mathcal C} &= \begin{bmatrix} -A_2 & -A_1 & B_2 & B_1 \end{bmatrix} = \begin{bmatrix}
0 & 0 & 0 & 0 & 1 & 0 \\
0 & 0 & 0 & 0 & 1 & 1 \end{bmatrix}.
\end{align*}
The system is not controllable (and thus, the treatment of \citep{Persis2019} does not work).
On the other hand, if we treat the system as two MISO systems, the order of each system is $\tilde n_1 = \tilde n_2 = 2$, and so, let $\chi_i = \text{col}(y_i(t-2),y_i(t-1),u(t-2),u(t-1)) \in \R^4$ for $i=1,2$.
The overall order is $2 \tilde n_1 + 2 \tilde n_2 = 8$, which is greater than 6 implying that we have non-minimal realization.
However, each system of $\chi_i$ is {\em controllable} so that identification of $\AAA_i$, $\BB_i$, and ${\mathcal C}_i$ for each output is enabled with the knowledge of $\tilde n_i$.
\end{rem}

\subsection{Implementation of D$^2$PC}\label{sec:4.3}

Based on Theorem \ref{thm:miso}, a new data-driven predictive control (D$^2$PC) for the plant \eqref{eq:sys0} can be proposed.
The first step for implementation is to choose $\bar n$ such that it is greater than or equal to the actual unknown order of the plant (Assumption \ref{ass:1prime}), and we assume that this is the case in this subsection.

With $\bar n$, construct $\bar \XX_{i,-}$ and $\bar \XX_{i,+}$ for each $i=1,\dots,p$, and $U_-$ to obtain $\bar \AAA_{d,i} \in \R^{2\bar n \times 2 \bar n}$ and $\bar \BB_{d,i} \in \R^{2 \bar n \times 1}$ by
\begin{align}\label{eq:AdBd}
	\begin{split}
		[ \bar {\mathcal A}_{d,i} \ \ \bar {\BB}_{d,i} ] &=  \bar \XX_{i,+} \begin{bmatrix} \bar \XX_{i,-} \\ U_- \end{bmatrix}^{\dagger}, \qquad i = 1,\dots, p.
	\end{split}
\end{align}
Let $\bar {\mathcal A}_{d} := {\rm block diag}(\bar {\mathcal A}_{d,1},\dots,\bar {\mathcal A}_{d,p})$ and $\bar {\mathcal B}_{d} := {\rm block diag}(\bar {\mathcal B}_{d,1},\dots,\bar {\mathcal B}_{d,p})$.
And let
$${\mathcal F} := \begin{bmatrix} \bar \AAA_d \\ \bar \AAA_d^2 \\  \vdots \\ \bar \AAA_d^N \end{bmatrix}, \
{\mathcal G} := \begin{bmatrix} \bar \BB_d & 0 &\dots  &0 \\ \bar \AAA_d \bar \BB_d & \bar \BB_d & \dots & 0 \\ \vdots \\ \bar \AAA_d^{N-1} \bar \BB_d & \bar \AAA_d^{N-2} \bar \BB_d & \dots & \bar \BB_d \end{bmatrix}$$
so that, with $\bar \chi(t) = {\rm col}(\bar \chi_1(t),\dots,\bar \chi_p(t)) \in \R^{2\bar n p}$, 
\begin{align*}
\begin{bmatrix} \bar \chi(t+1) \\ \bar \chi(t+2) \\  \vdots \\ \bar \chi(t+N) \end{bmatrix} = {\mathcal F} \bar \chi(t) + {\mathcal G} \begin{bmatrix} u(t) \\ u(t+1) \\ \vdots \\ u(t+N-1) \end{bmatrix}.
\end{align*}
Thus,
\begin{align*}
&\begin{bmatrix} y(t) \\ y(t+1) \\  \vdots \\ y(t+N-1) \end{bmatrix} = (I_{pN} \otimes e_{\bar n}^T) \begin{bmatrix} \bar \chi(t+1) \\ \bar \chi(t+2) \\  \vdots \\ \bar \chi(t+N) \end{bmatrix} \\
&= (I_{pN} \otimes e_{\bar n}^T) {\mathcal F} \bar \chi(t) + (I_{pN} \otimes e_{\bar n}^T) {\mathcal G} \begin{bmatrix} u(t) \\ u(t+1) \\ \vdots \\ u(t+N-1) \end{bmatrix}.
\end{align*}	
Then, the D$^2$PC algorithm is that, at time $t-1$, measure $y(t-1)$, construct $\bar \chi(t)$ with $u(t-1)$, solve 
\begin{align} \label{eq:DPC}
\begin{split}
\min_{\bar u} \ \ &\sum_{k=0}^{N-1} \left( \| \bar y_{k}-r(t+k)\|_Q^2 + \| \bar u_{k} \|_R^2 \right) \\
		\text{subject to}\ \ &  \bar y = (I_{pN} \otimes e_{\bar n}^T) {\mathcal F} \bar \chi(t) + (I_{pN} \otimes e_{\bar n}^T) {\mathcal G} \bar u \\
		&\bar u_k \in {\mathcal U},\ k=0, \dots, N-1, \\
		&\bar y_k \in {\mathcal Y},\ k=0, \dots, N-1,	
	\end{split}
\end{align}
where $\bar u = \text{col}(\bar u_0, \dots, \bar u_{N-1})$ and $\bar y = \text{col}(\bar y_0, \dots, \bar y_{N-1})$, and apply $u(t) = \bar u_0$ at time $t$.

Now, we present two recipes that make the proposed D$^2$PC less sensitive against measurement noise:
\begin{enumerate}

\item Collect multiple episodes of experiment data, compute $N_d$ multiple copies of \eqref{eq:AdBd}, get their average, and use them as $\bar {\mathcal A}_d$ and $\bar {\mathcal B}_d$.
The same idea of taking average may not be applied to the Hankel matrix of DeePC algorithm,
or to the matrices $\bar \XX_{i,-}$, $\bar \XX_{i,+}$ and $U_-$, unless the experiments are performed by the same input signals and the same initial conditions, because the averaging process not only reduces the level of the noise but also tends to reduce the level of the signals so that the signal-to-noise ratio remains the same.
On the contrary, the proposed averaging process is performed on the identified model $\bar {\mathcal A}_d$ and $\bar {\mathcal B}_d$, not on the raw input/output data, so that the aforementioned problem can be avoided.

\item Increase $\bar n$ (far beyond the estimated order of the plant).
It is observed that the closed-loop system is sensitive to the noise when $\bar n = n$, but simply by taking $\bar n$ a few more than $n$, the system becomes less sensitive to the noise.
We were not able to reasonably explain this phenomenon but will demonstrate it in the next section. Further study is called for.

\end{enumerate}

In the various benchmark examples of the next section, we treat $N_d$ and $\bar n$ as the design parameters of the proposed D$^2$PC, and demonstrate their effect.

\begin{figure}[!t]
\begin{center}
\includegraphics[width=0.48\textwidth]{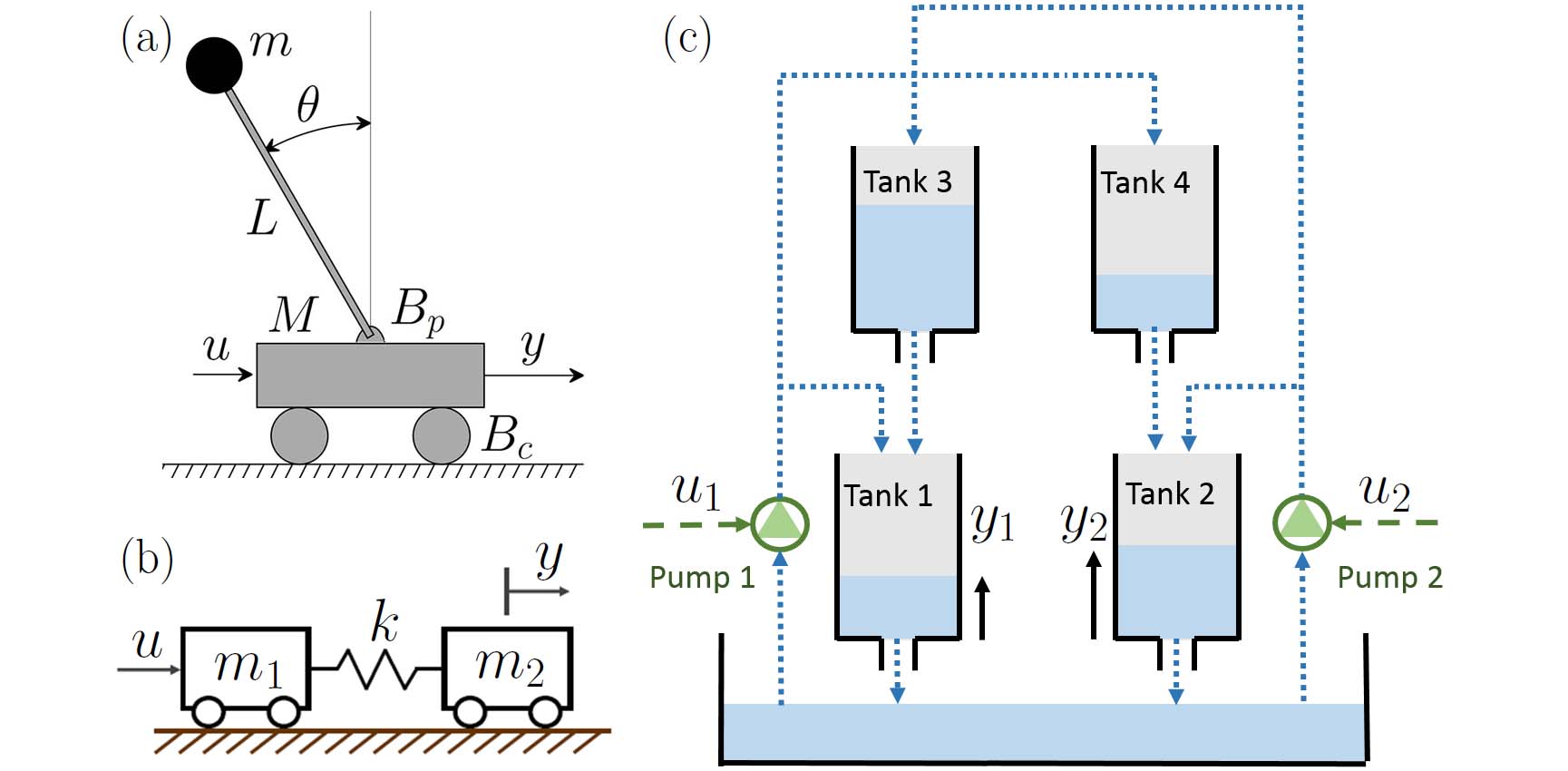}
\caption{Systems considered in Section \ref{sec:sim}}
\label{fig:1}
\end{center}
\end{figure}

%%%%%%%%%%%%%%%%%%%%%%%%%%%%%%%%%%%%%%%%%%%%%%%%%%%%%%%%%%%%%%%%%%%%%%%%%%%%%
\section{Simulation Study} \label{sec:sim}
%%%%%%%%%%%%%%%%%%%%%%%%%%%%%%%%%%%%%%%%%%%%%%%%%%%%%%%%%%%%%%%%%%%%

In this section, three benchmark examples are considered, and the proposed D$^2$PC and the DeePC (with/without the regularization technique) are compared.
As an optimization solver, the OSQP (Operator Splitting Quadratic Program) by \cite{OSQP} is employed to solve DeePC, rDeePC, and D$^2$PC.
The control performance is evaluated in terms of the mean absolute error (MAE), which is defined as 
\begin{equation*} 
	\text{MAE}= \frac{1}{N_{\sf sim}} \sum_{t=1}^{N_{\sf sim}} \|y(t)-y_{nom}(t)\|
\end{equation*} 
where $N_{\sf sim}$ is the simulation horizon, and $y_{nom}(t)$ is the output of the plant under the MPC based on the accurate model in the absence of measurement noise and the state estimate $\hat x(t)$ is set to $x(t)$ in \eqref{eq:MPC}.
Both $y_d$ (offline measurement) and $y$ (online measurement) are corrupted with additive random noise $n(t)$ with noise intensity $A_n$ (i.e. $\|n(t)\|_\infty \le A_n$).
Because the simulation outcome depends on the random noise, we carry out the same simulation 10 times to get the averaged value of MAE in this section.
The MATLAB codes used for the results in this section are available at \url{https://github.com/hyungbo/d2pc}.

%%%%%%%%%%%%%%%%%%%%%%%%%%
\subsection{Inverted Pendulum}

Consider the inverted pendulum in Fig.~\ref{fig:1}.(a), which has been widely used for the evaluation of newly designed control algorithm. 
The continuous-time system model and system parameters can be found in \citep{quanser_IP}.
Discretization with a sampling time of $0.1$ seconds yields
\begin{align*}
		A&= \begin{bmatrix}
		    1.208  &  0.106   &      0  &  0.096 \\
		    4.187  &  1.194   &      0  &  1.779 \\
		    -0.016 &  -0.001  &      1  &   0.070 \\
		    -0.299 &  -0.015  &      0  &   0.460
		\end{bmatrix}, \ B= \begin{bmatrix}
		 -0.022 \\  -0.414 \\  0.007 \\	 0.126
	\end{bmatrix}\\
		C&= [0 \ \ 0 \ \ 1 \ \ 0].
\end{align*}
Suppose that the cart should track a unit step signal $r(t)$ under the constraint that $-20 \le u(t) \le 20$, and the prediction horizon is chosen as $N=20$ with $Q=1000$ and $R=1$.

\begin{figure} [!t]
	\begin{center}
		\includegraphics [width=0.45\textwidth]{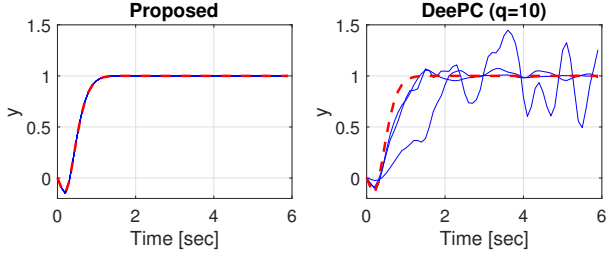}
		\caption{\label{fig:IP1} Inverted pendulum system ($A_n = 0$).
		Dotted red represents the nominal $y_{nom}(t)$, and solid blue represents the output $y(t)$ by D$^2$PC [left; 10 outputs are overlapped] and by DeePC [right; successful 3 results, out of 10, are drawn].}
	\end{center}
\end{figure}

We first consider the case where the system order is known exactly and there is no measurement noise.
For DeePC, the sampling input $u_d$ should be persistently exciting of order $(T_{\rm ini} + N + n = 28)$, and so, the minimum length $T$ of an episode is $(1+m)(T_{\rm ini}+N+n)-1 = 55$.
However, since the plant is unstable, the corresponding output $y_d$ is likely to grow unbounded. 
(In fact, our simulation yields $\|y_d(55)\|=6.565 \times 10^{10}$.)
With these data, our solver OSQP didn't work well, and so, we had to use the technique of multiple data set \citep{Waarde2020}, which yielded the length $T$ of each episode as $T > 28$.
On the other hand, D$^2$PC requires the persistent excitation of order $(2n+1 = 9)$ and the length $T$ of each episode to be $T \ge (1+m)(2n+1)-1 = 17$.
Simulation results are shown in Table \ref{tab:InvPend0}, in which the acronym FR stands for failure ratio of the optimization solver, and $q$ represents the number of data samples used for DeePC.
When failure of the solver occurs, it was not accounted for the computation of MAE.
Fig.~\ref{fig:IP1} compares the plant's output by D$^2$PC for $\bar n = 4$ and $N_d = 1$, and that by DeePC for $q=10$.
It is seen that the response by D$^2$PC is almost indistinguishable from the nominal trajectory.

Now, we consider the case where the system order is unknown but its upper bound is assumed to be $\bar n = 10$, and the plant's output is corrupted with measurement noise with $A_n = 10^{-4}$.
As a countermeasure against the noise, we used the regularized DeePC (rDeePC),
and took $N_d = 50$ for D$^2$PC.
The outcome is shown in Table \ref{tab:InvPend1}.

Finally, to see the effect of increasing $\bar n$, the simulations are carried out for various values of $\bar n$ while $N_d = 50$.
As seen from Table \ref{tab:InvPend2}, MAE of D$^2$PC tends to decrease as $\bar n$ increases.

\begin{table}
	\renewcommand{\arraystretch} {1.2}
	\caption{Inverted pendulum ($A_n=0$): $N_d = 1$ for D$^2$PC}
	\label{tab:InvPend0}
	\begin{minipage}{.73\textwidth}
		\begin{tabular}{c c c  c  c  c }
			\hline
			   & DeePC & DeePC & DeePC & DeePC & D$^2$PC   \\
				 & ($q$=1)  & ($q$=3) & ($q$=5) & ($q$=10) 
				 & ($\bar n=4$) \\
				\hline
				MAE & 0.890 & 0.869 & 0.498 & 0.146 &  $<0.001$ \\
			    FR & 0.9 & 0 & 0 & 0.7 & 0 \\
				\hline
		\end{tabular}
	\end{minipage}
\end{table}

\begin{table}
	\renewcommand{\arraystretch} {1.2}
	\caption{Inverted pendulum ($A_n=10^{-4}$): $N_d = 50$ for D$^2$PC}
	\label{tab:InvPend1}
	\begin{minipage}{.73\textwidth}
		\begin{tabular}{c c c  c  c  c }
			\hline
			   & DeePC & DeePC & rDeePC & rDeePC & D$^2$PC   \\
				 & ($q$=5)  & ($q$=10) & ($q$=5) & ($q$=10) 
				 & ($\bar n=10$) \\
				 \hline 
				MAE & N.A. & N.A. & N.A. & N.A. & 0.065  \\
				FR & 1 & 1 & 1 & 1 & 0 \\
				\hline
		\end{tabular}
	\end{minipage}
\end{table}

\begin{table}
	\renewcommand{\arraystretch} {1.2}
	\caption{Inverted pendulum ($A_n=10^{-4}$): Performance of D$^2$PC for various $\bar n$ with $N_d=50$}		
	\label{tab:InvPend2}
	\begin{minipage}{.77\textwidth}
		\begin{tabular}{c c c  c  c  c  c  c}
			\hline
			 $\bar n$  & 4  & 6 & 8 & 10 & 12 & 14   \\
			\hline
			MAE & N.A.    & 0.292 & 0.107 & 0.065 & 0.084 & 0.063 \\
			FR  & 1 & 0 & 0 & 0 & 0 & 0 \\
			\hline
		\end{tabular}
	\end{minipage}
\end{table}

%%%%%%%%%%%%%%%%%%%%%%%%%%
\subsection{Two-mass System}

As a second benchmark example, we consider two-mass system of Fig. \ref{fig:1}.(b), which has been widely used as a benchmark problem for robust controller design \citep{Wie1992acc}.
The parameters are assumed to be $m_1=1, m_2=0.1$, and $k=2$, which yields a marginally stable system because there is no friction.
The discrete-time model (using discretization with a sampling time of $T_s=0.1$) is given by
\begin{align*}
		A&= \begin{bmatrix}
		    0.990    &0.100   & 0.01   & 0.000 \\
		    -0.193   & 0.990  &  0.193  &  0.010 \\
		    0.098    &0.003   & 0.902   & 0.097 \\
		    1.928    &0.098   &-1.93   & 0.902	\\
		\end{bmatrix}, \ B= \begin{bmatrix}
		    0.005 \\    0.010 \\     0.000 \\    0.003
	\end{bmatrix} \\
	C&= [0 \ \ 0 \ \ 1 \ \ 0] .
\end{align*}
The control goal is to make $y(t)$ track a unit step signal $r(t)$ under the constraint that  $-2 \le u(t) \le 2$, and it is supposed that $N=20$, $Q=200$, and $R=1$.
Here, motivated by the observation in the previous subsection, let us take $\bar n = 20$ which is large enough compared to what is expected in practice for two-mass system.
Through various simulations, the regularization parameters for rDeePC are selected as $\lambda_g=500$ and $\lambda_y=5\times 10^5$.
For a fair comparison, $u_d$ of the same length $T=100$ is used for all methods.

\begin{figure} [!t]
	\begin{center}
		\includegraphics [width=0.45\textwidth]{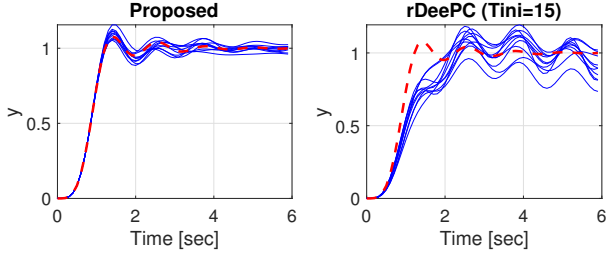}
		\caption{\label{fig:Cart_An001} Outputs of two-mass system with $A_n=0.01$.}
	\end{center}
\end{figure}

Simulation results for noise intensity $A_n=0.01$ are depicted in Fig. \ref{fig:Cart_An001}.
Table \ref{tab:TwoMass1} also shows the outcomes for various values of $A_n$.
(Failure ratio is not shown here since there were no failures.)

\begin{table}
	\renewcommand{\arraystretch} {1.2}
	\caption{Two-mass system: Comparison of MAE ($N_d = 1$)}
	\label{tab:TwoMass1}
	\begin{minipage}{.73\textwidth}
		\begin{tabular}{c c c  c  c  c }
			\hline
		   & DeePC & DeePC & rDeePC & rDeePC & D$^2$PC   \\
			$A_n$ & ($T_{\rm ini}$=4)  & ($T_{\rm ini}$=15) & ($T_{\rm ini}$=4) & ($T_{\rm ini}$=15) 
			 & ($\bar n=20$) \\
			\hline
			$10^{-8}$ & $<$0.001 & $<$0.001 & 0.397 & 0.093 & $<$0.001  \\
			$10^{-4}$ & 1.312 & 0.470 & 0.397 & 0.093 & $<$0.001 \\
			$10^{-2}$ & 0.993 & 1.523 & 0.486 & 0.092 & 0.009 \\
			$10^{-1}$ & 0.856 & 2.984 & 0.808 & 0.169 & 0.129 \\ 
			\hline
		\end{tabular} \\
	\end{minipage}
\end{table}

From Table \ref{tab:TwoMass2}, we again observe that MAE of D$^2$PC tends to decrease as $\bar n$ is increased, and from Table \ref{tab:TwoMass3}, it is seen that MAE of D$^2$PC tends to decrease with increasing $N_d$.
We also applied the same averaging technique to DeePC, that is, the computation of $U_p$, $U_f$, $Y_p$, and $Y_f$ are averaged over multiple data samples.
As seen in Table \ref{tab:TwoMass3}, it was not effective (as briefly discussed in Section~\ref{sec:4.3}).

\begin{table}
	\renewcommand{\arraystretch} {1.2}
	\caption{Two-mass system: MAE of D$^2$PC for various $\bar n$ ($N_d = 1$)}
	\label{tab:TwoMass2}
	\begin{minipage}{.77\textwidth}
			\begin{tabular}{c c  c  c c  c c}
				\hline
				 $\bar n$  & 4 & 6 & 8 & 10 & 15 & 20   \\
				\hline
				$A_n=10^{-2}$ & 4.951 & 0.842 & 0.237 & 0.057 & 0.012 & 0.009 \\
				$A_n=10^{-1}$ & 6.284 & 3.993 & 2.732 & 0.436 & 0.144 & 0.129 \\	
				\hline
			\end{tabular}
	\end{minipage}
\end{table}	
 
\begin{table}
	\renewcommand{\arraystretch} {1.2}
	\caption{Two-mass system ($A_n=0.1$): MAE for various $N_d$}
	\label{tab:TwoMass3}
	\begin{minipage}{.93\textwidth}
		\begin{tabular}{c c c  c  c  c c}
		\hline
		$N_d$  & 1 & 5 & 20 & 50 & 500   \\
		\hline
		D$^2$PC($\bar n=20$)      & 0.129 & 0.059 & 0.032 & 0.033 & 0.028 \\
		rDeePC($T_{\rm ini}$=15)  & 0.169 & 0.225 & 0.416 & 0.598 & 0.776 \\
		\hline
		\end{tabular}
	\end{minipage}
\end{table}

%%%%%%%%%%%%%%%%%%%%%%%%%%
\subsection{Four Tank System}

Our last example is a multi-input-multi-output (MIMO) system.
Consider a four tank system of Fig.~\ref{fig:1}.(c), whose discrete-time representation is given by \cite{Berberich2020}:
\begin{align*}
		A&= \begin{bmatrix}
		    0.921 & 0 & 0.041 & 0 \\
		    0 & 0.918 & 0 & 0.033 \\
		    0 & 0 & 0.924 & 0 \\
		    0 & 0 & 0 & 0.937
		\end{bmatrix}, \ B= \begin{bmatrix}
		    0.017 & 0.001 \\ 0.001 & 0.023\\
		    0 & 0.061\\ 0.072 & 0
	\end{bmatrix} \\
	C&= \begin{bmatrix}
		1 & 0 & 0 & 0 \\ 0 & 1 & 0 & 0
	\end{bmatrix}.
\end{align*}
As in \cite{Berberich2020}, the control goal is to make the output $y(t)$ track a setpoint $r(t)=[0.65, \ 0.77]^T$.
Most of design parameters are chosen as the same as in \cite{Berberich2020}: $N=30$, $Q=3I_2$, $R=0.01I_2$, and there are no input/output constraints. 
We also took the same parameters for rDeePC: $\lambda_g=0.1$, $\lambda_y=1000$, and $T=400$.

For actual plant, it is assumed that nothing is known but we assume that the order is less than $\bar n = 30$ (which is again far beyond the usual expectation of four tank system).
Computer simulations are carried out and the results are summarized in Table \ref{tab:4Tank1}.
From Tables \ref{tab:4Tank2} and \ref{tab:4Tank3}, we observe the same tendency as before for the MIMO system.

From the repeated simulation study, we found that a suitable choice of the regularization parameters for rDeePC is not trivial, but for D$^2$PC, choosing two parameters $\bar n$ and $N_d$ was relatively straightforward.

\section{Conclusion}

We have presented a new data-driven, output-feedback predictive control scheme for multi-input-multi-output, unknown, linear time-invariant plants.
The order of the plant need not be known, which is in a sharp contrast to other popular methods such as \citep{Lewis2012,Rizvi2018,Persis2019}.
There are only two tuning parameters $\bar n$ and $N_d$ for the proposed controller, and it was demonstrated through three benchmark examples that increasing both parameters makes the closed-loop less sensitive to the measurement noise.
Requiring relatively small length of episode data, it can be an effective method for unstable plants.
If there is no input/output constraint in the optimization problem, the QP of D$^2$PC
%in \eqref{eq:DPC} 
is analytically solved and the optimal control becomes a linear feedback.
Therefore, the proposed method can be considered as a constructive way to obtain a data-driven output-feedback LQR controller.

\begin{table}
	\renewcommand{\arraystretch} {1.2}
	\caption{Four tank system: Comparison of MAE ($N_d = 1$)}
	\label{tab:4Tank1}
	\hspace{-4mm}
	\begin{minipage}{.73\textwidth}
		\begin{tabular}{c c c  c  c  c }
			\hline
		   & DeePC & DeePC & rDeePC & rDeePC & D$^2$PC   \\
			$A_n$ & ($T_{\rm ini}$=4)  & ($T_{\rm ini}$=30) & ($T_{\rm ini}$=4) & ($T_{\rm ini}$=30)  & ($\bar n=30$) \\
			\hline
			$10^{-7}$ & $<$0.001 & $<$0.001 & 0.010 & 0.010 & $<$0.001 \\
			$10^{-3}$ & 0.952 & 0.939 & 0.013 & 0.010 & 0.001 \\
			$10^{-2}$ & 0.952 & 0.952 & 0.089 & 0.021 & 0.007 \\
			$10^{-1}$ & 0.952 & 0.952 & 0.515 & 0.200 & 0.074 \\  
			\hline
		\end{tabular}
	\end{minipage}
\end{table}
			
\begin{table}
	\renewcommand{\arraystretch} {1.2}
	\caption{Four tank system: MAE of D$^2$PC for various $\bar n$ ($N_d = 1$)}	
	\label{tab:4Tank2}
	\begin{minipage}{.73\textwidth}
		\begin{tabular}{c c  c  c  c  c c}
			\hline
			 $\bar n$  & 4 & 6 & 10 & 15 & 20 & 30   \\
			\hline
			$A_n=10^{-2}$ & 0.053 & 0.029 & 0.014 & 0.008 & 0.006 & 0.007 \\
			$A_n=10^{-1}$ & 0.660 & 0.408  & 0.189 & 0.096 & 0.079 & 0.074 \\
			\hline
		\end{tabular}
	\end{minipage}
\end{table}

\begin{table}
	\renewcommand{\arraystretch} {1.2}
	\caption{Four tank system ($A_n=0.1$): MAE for various $N_d$}
	\label{tab:4Tank3}
	\begin{minipage}{.73\textwidth}
		\begin{tabular}{c c c  c  c  c c c}
			\hline
			$N_d$  & 1 & 5 & 20 & 50 & 500   \\
			\hline
			D$^2$PC($\bar n=30$)     & 0.074 & 0.033 & 0.020 & 0.015 & 0.013 \\
			rDeePC($T_{\rm ini}=30$) & 0.200 & 0.122 & 0.124 & 0.184 & 0.418 \\	
			\hline
		\end{tabular}
	\end{minipage}
\end{table}

\end{document}